\documentclass{Interspeech2024}
\usepackage{hyperref}
\usepackage{cite}
\hypersetup{hypertex=true,
            colorlinks=true,
            linkcolor=blue,
            anchorcolor=blue,
            citecolor=blue}
\usepackage{amsmath}
\usepackage{tabularray}

\newcommand\blfootnote[1]{%
\begingroup
\renewcommand\thefootnote{}
\footnote{#1}%
\addtocounter{footnote}{-1}%
\endgroup
}

\interspeechcameraready

\title{Improving Speech Enhancement by Integrating Inter-Channel and Band Features with Dual-branch Conformer}

\name[affiliation={1}]{Jizhen}{Li}
\name[affiliation={1}]{Xinmeng}{Xu}
\name[affiliation={1, 2, \star}]{Weiping}{Tu}
\name[affiliation={1, 2}]{Yuhong}{Yang}
\name[affiliation={1}]{Rong}{Zhu}

\address{
  $^1$NERCMS, School of Computer Science, Hubei Luojia Laboratory, Wuhan University, China\\
  $^2$Hubei Key Laboratory of Multimedia and Network Communication Engineering, Wuhan University, China}
\email{tuweiping@whu.edu.cn}


\begin{document}

\maketitle

\begin{abstract}
    Recent speech enhancement methods based on convolutional neural networks (CNNs) and transformer have been demonstrated to efficaciously capture time-frequency (T-F) information on spectrogram. However, the correlation of each channels of speech features is failed to explore. Theoretically, each channel map of speech features obtained by different convolution kernels contains information with different scales demonstrating strong correlations. To fill this gap, we propose a novel dual-branch architecture named channel-aware dual-branch conformer (CADB-Conformer), which effectively explores the long range time and frequency correlations among different channels, respectively, to extract channel relation aware time-frequency information. Ablation studies conducted on DNS-Challenge 2020 dataset demonstrate the importance of channel feature leveraging while showing the significance of channel relation aware T-F information for speech enhancement. Extensive experiments also show that the proposed model achieves superior performance than recent methods with an attractive computational costs.
\end{abstract}

\begin{figure*}
    \centering
    \includegraphics[width=0.9\linewidth]{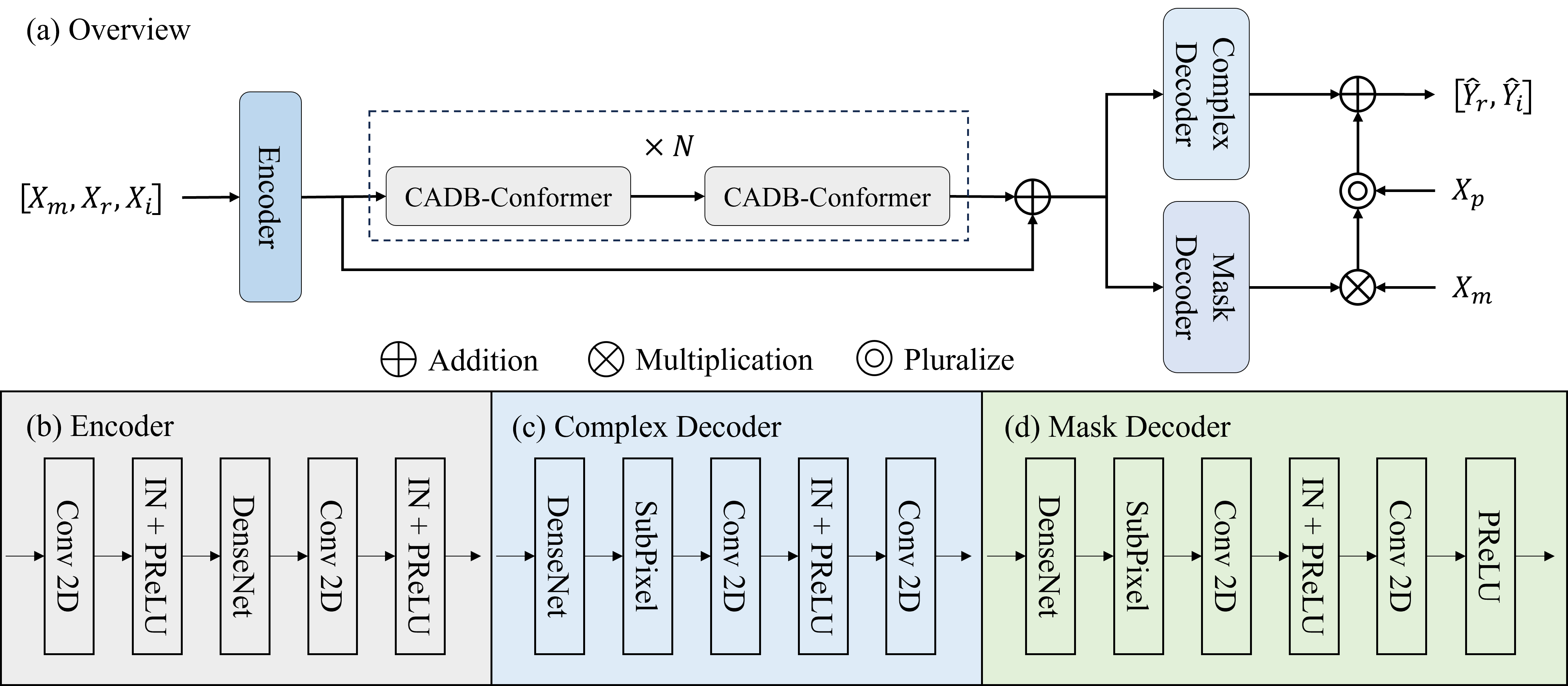}
    \caption{Overall enhancement process of the proposed CADB-Conformer. (a) An overview of the proposed CADB-Conformer architecture. (b) The encoder architecture of CADB-Conformer. (c) Complex decoder unit. (d) Mask decoder unit.}
    \label{fig:overview}
\end{figure*}

\section{Introduction}

Speech enhancement (SE) aims at separating clean speech from acoustically challenging environments with various disturbances and improving the perceptibility and quality of the speech signal \cite{benesty2006speech}. SE technologies find widespread applications in various domains, including telecommunications, hearing aids, automatic speech recognition (ASR), and audio broadcasting \cite{cao2022cmgan}. \blfootnote{$^\star$Corresponding author.}

Recently, the data-driven approaches based on Deep Neural Networks (DNNs) have been extensively investigated and can be categorized into time-domain methods \cite{stoller2018wave,luo2019conv,defossez2020real} and time-frequency (T-F) domain methods \cite{zhao2018convolutional,yin2020phasen,wang2022tf, xu2024adaptive}. Despite the intuitive and low-latency characteristics of time-domain methods, their performance is limited in complex noise environments due to the lack of geometric structure \cite{hao2021fullsubnet}. Consequently, T-F methods remain the mainstream in the realm of speech enhancement. T-F domain methods take the T-F representation of the original signal as input, such as the short-time Fourier Transform (STFT), and directly estimate the complex spectrum of clean speech in an end-to-end manner \cite{opochinsky2024single}. Furthermore, complex methods \cite{zhao2021monaural,hu2020dccrn} are designed to handle phase-related issues more effectively in the complex domain. The CNN-based module maps features to a high-dimensional space using diverse kernels, where the global context information in the T and F domains varies across channels. Despite the excellent performance of CNNs on extracting local T-F features, convolution operation treats the spectral and channel-wise features equally, which makes it challenging to perform fine-grained processing on the temporal, frequency, and channel dimensions. 

Recent works represented by dual-path models like DPRNN \cite{luo2020dual} and DPCRN \cite{le2021dpcrn}, process the intra/temporal bands and inter/spectral bands of the feature separately, effectively modeling both domains and exhibiting state-of-the-art performance in SE. Subsequently, transformer \cite{vaswani2017attention} architectures are applied to the dual-path model \cite{subakan2021attention,lin2022dptnet,yu2022dual,wang2021tstnn,dang2022dpt}, replacing long short-term memory (LSTM) for improved modeling capabilities. Specifically, the DPTNet \cite{lin2022dptnet} introduces direct context-aware modeling into speech separation for the first time. The incorporation of self-attention enables direct interaction among elements in the speech sequence, facilitating information transmission. Moreover, the integration of recurrent neural networks (RNNs) into the original transformer allows it to learn the sequential information of the speech sequence without the need for position encoding. However, it greatly increase the model complexity. Despite the success of CNNs and dual-path architecture in sequence modeling, it fails to consider the inter-channel feature correlation and overlooks the learning of differentiated information at different scales, consequently hindering the representational capacity of deep networks.

Therefore, it is highly necessary to design a dedicated module in channel dimension to facilitate high-level alignment across different channel features. Motivated by previous research, we propose a dual-branch architecture, termed channel-aware dual-branch conformer (CADB-Conformer). Conformer \cite{gulati2020conformer} model combines characteristics of CNNs and transformer, enabling comprehensive modeling of both global and local information with relatively low complexity. In this work, in addition to the band feature branch (BFB) employed for fine-grained extraction of T and F domain features, we have meticulously designed the channel feature branch (CFB) to adaptively rescale each channel-wise feature by modeling the interdependencies across feature channels. CFB allows the model to concentrate on more useful channels and enhance discriminative learning ability. The contributions of this work are threefold:

\begin{itemize}
\item We investigate CFB to capture the features of channel dimension and efficiently integrate with the network backbone. Ablation studies demonstrate that channel features can significantly improve the network performance.

\item Our proposed model proficiently extracts the T and F information across channels, enabling better integration of information from different channels in a long-range manner.

\item Experiments perform on DNS-Challenge 2020 datasets \cite{reddy2020interspeech} show that our proposed CFB module achieves 0.07, 0.45 and 0.84 improvements in PESQ, STOI, and SDRi with the SNR set to 0dB.
\end{itemize}

\section{Methodology}

In this section, we provide a detailed description of the proposed CADB-Conformer neural network, as shown in Figure \ref{fig:overview}, which efficiently integrates 2D T-F features among channels and 1D band features extracted by the two branch.

\begin{figure*}
    \centering
    \includegraphics[width=0.9\linewidth]{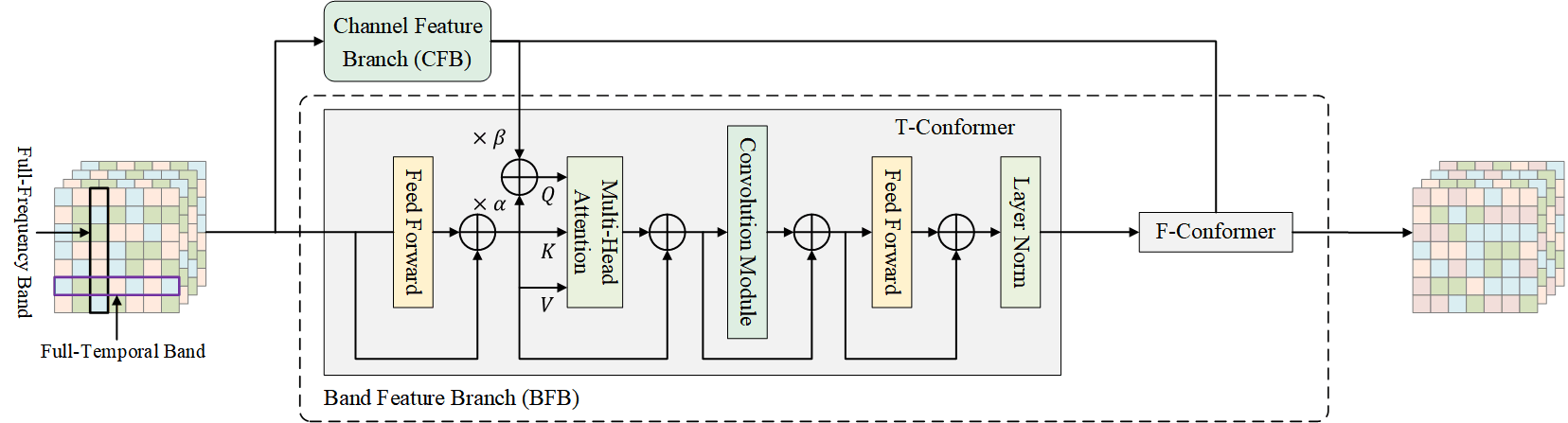}
    \caption{The detail of proposed CADB-Conformer module, which contains Channel Feature Branch (CFB) and Band Feature Branch (BFB).}
    \label{fig:CADB-Conformer}
\end{figure*}

\begin{figure}[t]
    \centering
    \includegraphics[width=\linewidth]{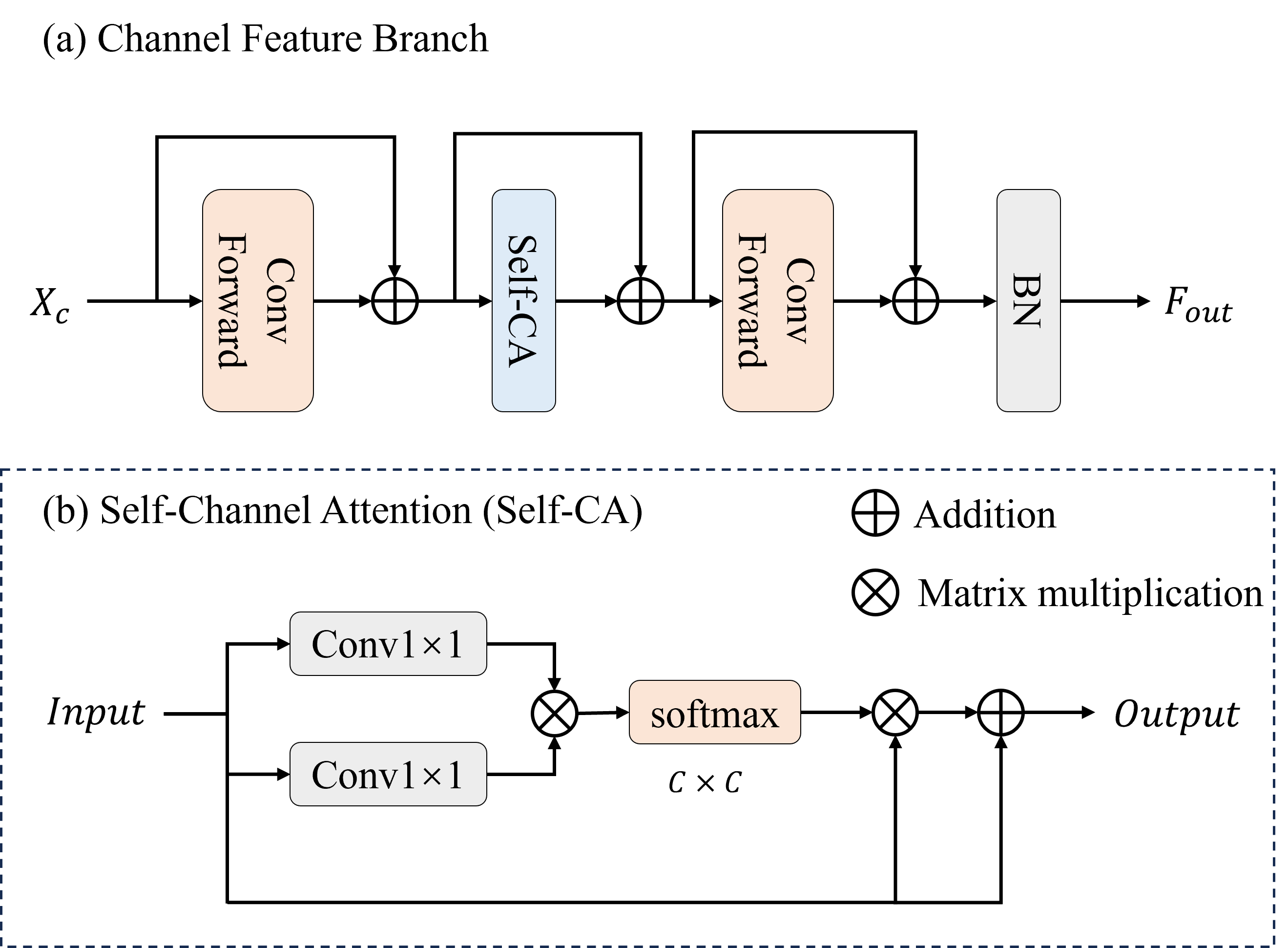}
    \caption{(a) The architecture of Channel Feature Branch. (b) The detail of Self-Channel Attention (Self-CA).}
    \label{fig:Channel Feature Branch}
\end{figure}

\subsection{Overview}

For a noisy speech waveform $ x\in \mathbb{R} ^{L\times 1} $, its real part spectrum and imaginary part spectrum, denoted as $X_{r} \in \mathbb{R} ^{T\times F\times 1} $ and $X_{i} \in \mathbb{R} ^{T\times F\times 1} $ respectively, are obtained through STFT, where T and F represent the time and frequency dimensions respectively. Following the approach proposed by \cite{cao2022cmgan}, the compressed spectrogram $X$ is obtained through power-law compression:
\begin{equation}
    X= X_{r} +jX_{i} = X_{m} e^{jX_{p}},
\end{equation}
where $X_{m}$ and $X_{p}$ denote the magnitude and phase components, respectively. To fully leverage the information within the time-frequency spectrum, the proposed CADB-Conformer takes an input $X$, which is constructed by concatenating the magnitude and the complex spectrum.

The proposed encoder-decoder architecture comprises one encoder and two decoders, each of the three module contains a dilated DenseNet, which contains four convolution blocks with dense connections, the dilation factors of each block are set to {1, 2, 4, 8}. The input $X\in \mathbb{R} ^{B\times T\times F\times 3} $ undergoes encoding to obtain a high-dimensional feature representation, where $B$ represents the batch size. The encoder's architecture, illustrated in Figure \ref{fig:overview}(b), input $X$ is preprocessed and the frequency dimension is compressed to half of the original size for efficient calculations. The two decoders individually predict the complex components and magnitude mask to reconstruct the target signal. The decoder structures resemble the encoder, as depicted in Figure \ref{fig:overview}(c) and (d). The complex decoder predicts the complex spectrum $\hat{Y} _{r}^{c} $ and $\hat{Y}_{i}^{c} $ of the target speech in a generative manner, while the magnitude decoder predicts the magnitude mask $\hat{Y}_{m}$. The target speech signal $\hat{y}$ predicts by CADB-Conformer is designed as:
\begin{equation}
\begin{cases}\hat{Y}_{r}= \hat{Y} _{r}^{c} +  \hat{Y} _{m} \cos X_{p}, 
    \\\hat{Y}_{i}= \hat{Y} _{i}^{c} +  \hat{Y} _{m} \sin X_{p}, 
    \\\hat{y}=iSTFT(\hat{Y}_{r} + j\hat{Y}_{i}),
\end{cases}
\end{equation}
where $iSTFT$ represents the inverse Short-Time Fourier Transform.

\subsection{Channel-Aware Dual-Branch Conformer Module}

The design of dual-branch model based on the conformer architecture proposed in \cite{gulati2020conformer}, which integrates the potent modeling capability of the transformer for global information and the fine-grained feature processing ability of CNN for local features. The detail of the proposed CADB-Conformer is shown in Figure \ref{fig:CADB-Conformer}: (1) The band feature branch comprises two cascaded conformer modules, each specifically handling the temporal and frequency domain of the spectrogram. (2) The design of channel feature branch followed by conformer module, which directly facilitates interaction between the T and F domain on a 2D level and feature alignment in channel dimension. (3) To selectively utilize T-F information extracted by channel feature branch, we explore the attention mechanism in band feature branch for feature fusion.

\subsubsection{Channel Feature Branch}

To efficiently harness the T-F information inherent in the spectrogram and align features on the channel dimension, we meticulously designed the channel feature branch (CFB) similar to conformer module, as shown in Figure \ref{fig:Channel Feature Branch}(a). CFB adopts a sandwich-like structure inspired by \cite{xu2023pcnn}, wherein the high-dimensional representation $X_{c} \in \mathbb{R} ^{B\times T\times \check{F} \times 3} $ of the time-frequency spectrogram is entered into the ConvForward module to obtain an enriched feature representation, where $\check{F} = \frac{F}{2} $. To investigate the global context information in both the temporal and frequency domains among each channel, the representation is then fed into the Self-Channel Attention (Self-CA) module to capture the long-range dependencies among channels. 

Inspired by \cite{deng2024multi}, Self-CA unfolds the temporal and frequency dimensions of the feature to obtain a one-dimensional representation $F_{in} \in \mathbb{R} ^{N\times 1} $, where $N = T \times \check{F}$, as shown in Figure \ref{fig:Channel Feature Branch}(b). The computation of Self-CA is as follows:
\begin{equation}
\begin{cases}Q/K=F_{in} \odot (\circledS Conv(F_{in})),
    \\W=\circledS(Q^{\top }  \otimes K),
    \\F_{out}=F_{in} \odot W + F_{in},
\end{cases}
\end{equation}
where $Conv(\cdot )$ denotes convolution operation, $\circledS(\cdot )$ denotes Softmax operation, $\odot$ denotes element-wise multiplication, $\otimes$ denotes matrix multiplication, respectively. To facilitate a high level of alignment across different channels and reducing computational complexity, we generate weight vectors $W$ of size $C\times C$ instead of $N\times N$. Such Self-CA mechanism enhances the discriminative learning capability of the proposed network architecture towards inter-channel features by explicitly adjusting the weights of different channel features to model long-range dependencies among channels.

Following Self-CA, we also incorporate the ConvForward module for further feature processing. To facilitate the efficient propagation of informative signals, we have incorporated skip connections into each module of the CFB .It is noteworthy that we do not add the convolutional module proposed by the conformer after the self-channel attention mechanism. Because the results of the CFB will be utilized in the attention weight calculation of the Band Feature Branch, where local feature extraction will be conducted.

\subsubsection{Band Feature Branch}

Inspired by \cite{chae2023exploiting}, the band feature branch (BFB) comprises two cascaded conformer modules, each dedicated to feature extraction along the time and frequency dimensions of the spectrogram. As shown in Figure \ref{fig:CADB-Conformer}, to utilize the T-F information extracted by CFB while mitigating the redundancy caused by global interactions between the time and frequency dimensions, we explore the attention mechanism within the conformer architecture. Specifically, for each conformer block, the output $X_{f}$ of the first FeedForward module and the output $F_{out}$ of the CFB are weighted and combined to form the query $Q$ for the attention mechanism. The calculation of $Q, K, V$ in the attention mechanism of the BFB is as follows:
\begin{equation}
\begin{cases}\hat{Q} = \alpha X_{f} + \beta F_{out} , 
    \\Q = Linear(\hat{Q}) ,
    \\K/V = Linear(F_{out}),
\end{cases}
\end{equation}
where $\alpha$ and $\beta$ represent the weights of $X_{f}$ and $F_{out}$ respectively and $Linear$ 
denotes fully connected layer. Leveraging the selection capability of the attention mechanism for query enables the comprehensive utilization of both 2D global T-F information and 1D band information while minimizing redundancy in the 2D information. Subsequently, the band feature obtained through the attention mechanism are fed into convolutional modules to extract local features. By employing attention mechanism instead of directly summing the features, the network is able to learn the relationship between global band features and channel-aligned features, selectively enhancing the representational ability of the network by utilizing useful information. Furthermore, in order to reduce the computational burden of the neural network, the T-Conformer and F-Conformer within each CADB-Conformer module share the same CFB.

\begin{table*}
\centering
\caption{Comparisons of baseline models on DNS-Challenge 2020 datasets in terms of PESQ, STOI, and SDRi}
\begin{tblr}{
  column{even} = {c},
  column{3} = {c},
  column{5} = {c},
  column{7} = {c},
  column{9} = {c},
  column{11} = {c},
  cell{1}{11} = {r=2}{},
  vline{2,5,8,11} = {-}{},
  hline{1,3-4,8-9} = {-}{},
  hline{2} = {1-10}{},
}
SNR          &      & -5 dB     &       &      & 0 dB      &       &      & 5 dB      &       & Param. \\
Metric       & PESQ & STOI (\%) & SDRi  & PESQ & STOI (\%) & SDRi  & PESQ & STOI (\%) & SDRi  &        \\
Unprocess    & 0.56 & 67.00     & -     & 0.75 & 76.39     & -     & 0.95 & 83.30     & -     & -      \\
Conv-TasNet \cite{luo2019conv}  & 1.70 & 83.07     & 15.13 & 2.28 & 88.38     & 11.00 & 2.36 & 91.28     & 10.75 & 3.5M   \\
RUI-NSNet \cite{cao2023refining}    & 1.73 & 83.00     & 15.11 & 2.36 & 89.03     & 15.67 & 2.50 & 91.74     & 11.62 & 4.0M   \\
DPRNN \cite{luo2020dual}        & 1.90 & 85.77     & 16.53 & 2.52 & 89.90     & 16.15 & 2.58 & 92.93     & 11.76 & 2.6M   \\
DPTNet \cite{lin2022dptnet}       & 1.90 & 87.59     & 15.90 & 2.50 & 90.50     & 16.04 & 2.64 & 93.66     & 11.71 & 2.8M   \\
CADB-Conformer & 2.17 & 88.52     & 17.37 & 2.66 & 91.11     & 16.65 & 2.86 & 94.70     & 12.59 & 2.0M   
\end{tblr}
\end{table*}

\begin{table}
\centering
\caption{Results of the ablation study on DNS-Challenge 2020 datasets with SNR set to 0dB}
\begin{tblr}{
  column{even} = {c},
  column{3} = {c},
  column{5} = {c},
  vline{2,5} = {-}{},
  hline{1-3,7} = {-}{},
}
Metric          & PESQ & STOI (\%) & SDRi & Param. \\
CADB-Conformer    & 2.66    & 91.11         & 16.65      & 2.0M   \\
w/o CFB         & 2.59    & 90.66         & 15.81      & 1.8M   \\
w/o F-Conformer & 2.33    & 88.18         & 15.23      & 1.5M   \\
w/o T-Conformer & 1.95    & 85.71         & 10.53      & 1.5M   \\ 
w/o BFB         & 1.79    & 82.72         & 11.24      & 1.0M
\end{tblr}
\end{table}

\section{Experiments}

\subsection{Datasets}

We trained the CADB-Conformer on the Interspeech 2020 DNS challenge datasets \cite{reddy2020interspeech}, which consists of over 60,000 clean speech and noise clips sampled at 16 kHz. We randomly selected 25,000 samples from the clean and noise dataset for mixing. The SNR range of the mixtures is set between -5 and 20 dB at 5 intervals, with speech segment lengths set to 4 seconds. Among these, 23,000 clips were chosen for training randomly and another 2,000 clips were allocated for validation. In addition, we utilize an equal number of clean speech samples and noise samples, without any repetition or reuse of noise instances, which aims to improve the generalization ability of the model.

\subsection{Experimental setup}

The input of the model is the complex spectrum obtained by STFT of the original signal with window-size set to 400. The resulting high-dimensional features obtained by the final encoder consist of 101 frequency bands. The repetition count N of CADB-Conformer is set to 4. Based on our experimental findings, we set $\alpha$ and $\beta$ both to 0.5. The experiment adopts scale-invariant SNR (SI-SNR) \cite{le2019sdr} as the training target with the optimizer selected as Adam \cite{kingma2014adam}. In addition, we utilizing a decaying learning rate initially set to 0.001.

\section{Results and discussion}

\subsection{Baselines and evaluation metrics}

In order to test the performance under various unknown noise, we compared our CADB-Conformer with several baseline models, such as Conv-TasNet \cite{luo2019conv}, DPRNN \cite{luo2020dual}, DPTNet \cite{lin2022dptnet}and RUI-NSNet \cite{cao2023refining}. Under the same training target, we evaluated the performance of these models on the test sets of DNS-Challenge 2020 with SNR of -5, 0 and 5 dB.

We employed three objective evaluation metrics: perceptual evaluation of speech quality (PESQ) \cite{rix2001perceptual}, short-time objective intelligibility (STOI) \cite{taal2010short}, and Signal-to-Distortion Ratio Improvement (SDRi) \cite{le2019sdr}, to assess the performance of the proposed model. Comparisons were made with baseline models on different SNR test sets, as shown in Table 1. It can be observed that our proposed CADB-Conformer neural network achieves superior performance compared to the baseline models. Compared with the representative dual-path architecture DPTNet, CADB-Conformer was tested on DNS-Challenge 2020 datasets with an SNR range of -5 to 5, achieving an average improvement of 0.22, 0.86 and 0.99 in PESQ, STOI and SDRi, respectively. Moreover, our proposed model utilizes minimal parameters, occupying only 2.0M.

\subsection{Ablation study}

To validate the effectiveness of our proposed modules, we conducted ablation experiments on DNS-Challenge 2020 datasets with SNR set to 0 dB. We also utilize PESQ, STOI, and SDRi as evaluation metrics, as shown in Table 2. We conducted ablation experiments by removing the CFB module (w/o CFB) to validate that incorporating channel features can effectively improve the quality of speech enhancement. The ablation experiments showed that modeling the long-range temporal and frequency domain relationships between channels by CFB can improve the performance of model with improvements of 0.07, 0.45, and 0.84 in PESQ, STOI, and SDRi, respectively. Remarkably, Adding the CFB module only increased the total number of model parameters by 0.2M

The proposed CADB-Conformer architecture incorporates Temporal and Frequency Conformer architectures as its backbone. We demonstrate the contributions of each module in the BFB by removing T-Conformer (w/o T-Conformer), F-Conformer (w/o F-Conformer) and the entire band feature branch (w/o BFB) separately. 

\section{Conclusions}

In this work, we propose CADB-Conformer neural network for single-channel speech enhancement in the time-frequency domain. Our model combining the robust handling capabilities of band information by the dual-path network and facilitating a high level of alignment across different channels. Experimental results demonstrate the lightweight nature of our proposed model and its superior performance compared to some baseline models on the Interspeech 2020 DNS challenge datasets. Additionally, ablation experiments highlight the effectiveness of the proposed channel feature branch. In the future, we will train and evaluate the transferability of our model in more complex acoustic environments, such as those involving reverberation.
\section{Acknowledgement}
This work was supported in party by the National Nature Science Foundation of China (No.62071342, No.62171326), the Special Fund of Hubei Luojia Laboratory (No.220100019), the Hubei Province Technological Innovation Major Project(No.2021BAA034) and the Fundamental Research Funds for the Central Universities (No.2042023kf1033).

\bibliographystyle{IEEEtran}
\bibliography{mybib}

\end{document}